\documentclass[twocolumn,amsmath,amssymb,prb]{revtex4}

\usepackage{amsmath}
\usepackage{graphicx}

\begin{document}
\title{Phase diagram of a $d$-wave superconductor
with Anderson impurities
}

\author{L.~S.~Borkowski}
\affiliation{Faculty of Physics,
Adam Mickiewicz University, Umultowska 85, 61-614 Poznan, Poland}

\begin{abstract}
We present a self-consistent solution for a model of a
$d$-wave superconductor with finite concentration
of Anderson impurities at zero temperature
using the slave boson method.
We show how the phase diagram depends
on the strength of interaction between
impurity and extended states.
For fixed impurity level energy $E_0$ in the Kondo limit
there is one superconducting-normal
state transition for all impurity concentrations $n$.
When $E_0$ is close the Fermi energy
there are three such transitions for impurity
concentration exceeding certain minimum value $n(\Gamma_0,E_0)$.
If hybridization $\Gamma_0$ is fixed and the depth of the impurity level
is varied, there are always two transitions for concentration
above $n(\Gamma_0,E_0)$.
\end{abstract}


\maketitle


The problem of magnetic and nonmagnetic impurities
in superconductors (SC) has a long history.
Recent years brought new experimental results, most of which
concern strongly correlated systems such as high-temperature
and heavy-fermion superconductors.\cite{pjh08}
Many of these studies are complicated
by the uncertainty about the symmetry
of the order parameter in various compounds
as well as the precise nature of the interaction
between localized and extended states.
The proximity to antiferromagnetism in many of these materials
poses additional challenges.\\
\indent
Scanning tunneling microscopy
experiments in $\rm{Bi_2Sr_2CaCu_2O_{8+\delta}}$
reveal inhomogeneities on a nanoscopic scale.\cite{cren2000,
davis2001,howald2001,davis2002,davis2005}
Local variations of chemical composition, e.g.
presence of excess oxygen atoms and cationic
disorder outside the $\rm{CuO_2}$ planes
might be responsible for these modulations
of the local gap and local density of states (DOS).
In bismuth strontium copper oxide (BSCCO)
doped with magnetic impurity Ni
resonances in local density of states
attributed to impurity
were observed only in small-gap
domains and no such resonances were
seen in large-gap regions.\cite{davis2002}
This sensitivity of local DOS to the size of the gap
may result from the proximity to the critical point
where magnetic impurities become decoupled from
the superconductor.\cite{wf90,lsb92,ingersent98,vojta2003}\\
\indent
Motivated by these works we will try to understand
how the phase diagram of a superconductor with
magnetic impurities
depends on the position of the impurity level,
the strength of interaction between localized state
and conduction electrons and the impurity concentration.
By changing the depth of the impurity level we
can switch from the Kondo limit
to the mixed valence regime. The phase diagram
in these two cases is qualitatively different.
Knowing the general form of the phase diagram
may improve our understanding of experiments in
impurity-doped superconductors.

\tighten
\section{Model}
The model consists of electrons in a conduction band~with
BCS-type pairing interacting with Anderson impurity,

\begin{equation}
\begin{split}
H=&\sum_{k,m}\epsilon_k c^\dagger_{km}c_{km}
+E_0\sum_m f^\dagger_m f_m
+V\sum_{k,m}[c^\dagger_{k,m}f_mb+H.c.]\\
&+\sum_{k,m}[\Delta(k)c^\dagger_{km}c^\dagger_{-k-m}+H.c.].
\end{split}
\end{equation}
The operator $c^\dagger_{km}$ creates electron in a spin-orbit
partial wave state of angular momentum $m$ and momentum $k$.
The energies $\epsilon_k$ lie in a band of half-width $D$
and constant density of states
$N_0 = 1/2D$. The impurity state has energy $E_0$
and its hybridization matrix element with extended states is $V$.
The constraint $n_f+b^\dagger b = 1$ is added
to prevent double occupancy of the impurity site.

We assume a two-dimensional $d$-wave order parameter of the form
$\Delta(k) = \Delta_0 \cos(2\phi)$, where $\phi$ is the angle
in the $k_x-k_y$ plane.
In the mean field approximation the dynamics of the boson fields
is neglected and the boson fields are replaced by their
expectation values $<b^\dagger>=<b>=z^{1/2}$.
Minimizing the free energy
with respect to the auxiliary boson field and position of the
many-body resonance $\epsilon_f$
we arrive at the mean field equations

\begin{equation}
\label{MF1}
\frac{1}{N} = - {\textrm Im}\int_{-\infty}^\infty
d\omega f(\omega) \frac{1}{2} 
\textrm{Tr} (\tau_0+\tau_3) \textbf{G}_f (\omega+i0^+) \quad ,
\end{equation}

\begin{equation}
\label{MF2}
\begin{split}
{\frac{E_0-\epsilon_f}{V^2}}
= {\rm Im} \int_{-\infty}^\infty d\omega f(\omega)
\frac{1}{2} \textrm{Tr}
[(\tau_0+\tau_3)\\
\times \textbf{G}^0 (\omega + i0^+)
\textbf{G}_f (\omega + i0^+)] \quad ,
\end{split}
\end{equation}
where the conduction electron Green's function in clean superconductor
${\rm \bf G}^0$ is given by

\begin{equation}
\left(\textbf{G}^0(\omega)\right)^{-1} =
\omega \tau_0 - \epsilon_k \tau_3 - \Delta(k)( \tau_1-i\tau_2) \quad ,
\end{equation}

and the full impurity Green's function is

\begin{equation}
\label{Gf}
\textbf{G}_f^{-1}(\omega) =
\omega \tau_0 - \epsilon_f \tau_3 - \Sigma_f(\omega) \quad .
\end{equation}

\begin{figure}[t!p]
\includegraphics[width=0.9\columnwidth]{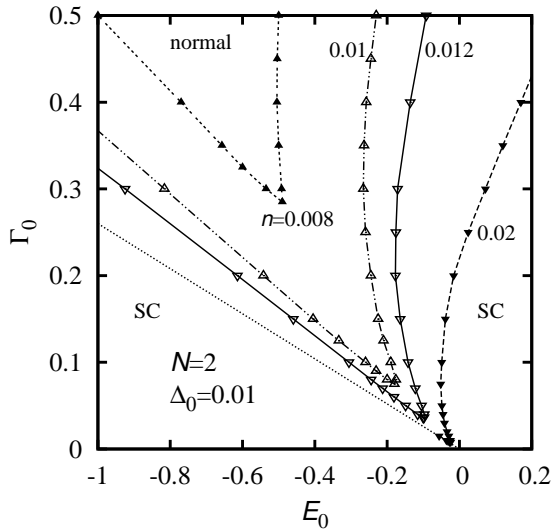}
\vskip -2.3cm
\caption{The phase diagram of $d$-wave superconductor with Anderson
impurities for four values of impurity concentration.
The lines separate the normal state from the superconductor (SC).
The broken line shows location of the impurity quantum phase
transition. Below that line the impurities
are decoupled from the superconductor. The slope of this line
is approximately -0.26 which agrees
with the numerical renormalization
group\cite{ingersent98} result and a single-impurity large-$N$
calculation for a d-wave
superconductor.\cite{yu2002}
}
\label{fig1}
\end{figure}

Conduction electron Green's function
is averaged over impurity positions in the usual way.
We have to solve the system of equations (\ref{MF1}), (\ref{MF2}),
together with the gap equation

\begin{equation}
\Delta(k) = \int_{-\infty}^{\infty} d\omega
f(\omega) \sum_{k^\prime} V_{kk^\prime}
{\textrm Tr} \frac{1}{2}{(\tau_1-i\tau_2)
\textbf{G}(k^\prime,\omega)}.
\end{equation}

For each frequency $\omega$ we solve self-consistently the Dyson
equations for conduction electron and impurity self-energies

\begin{equation}
\Sigma(\omega) = \textbf{G}^0(\omega) - \frac{nN}{2\pi N_0}
\Gamma \textbf{G}_f(\omega),
\end{equation}

\begin{equation}
\Sigma_f(\omega) = \textbf{G}_f^0(\omega)
- \Gamma \sum_k \textbf{G}(k,\omega),
\end{equation}
where $n$ is the concentration of impurities, $N$ is the degeneracy
of the impurity energy level $E_0$, $\Gamma=z\Gamma_0=z\pi N_0 V^2$.
Here we assumed $N=2$.

The self-energy $\Sigma_f(\omega)$
in eqn. (\ref{Gf}) in general contains both diagonal and off-diagonal
terms in particle-hole space. Here the off-diagonal term
is zero due to vanishing of the Fermi surface average of
the off-diagonal part of the conduction electron Green's function.

The renormalized frequencies are solutions of the following equations

\begin{equation}
\bar{\omega} = \omega + \Gamma
\langle \frac{\widetilde{\omega}}{\left( \Delta^2(k)
- {\widetilde{\omega}}^2\right)^{1/2}}\rangle ,
\end{equation}
and

\begin{equation}
\widetilde{\omega} = \omega + {\frac{nN}{2\pi N_0}}
\frac{\bar{\omega}}{(-\bar{\omega}^2+\epsilon^2_f)} .
\end{equation}
where brackets denote average over the Fermi surface.
The presence of full Green's functions under integrals
of eqs. (3), (4) and (7) means eqs. (10) and (11)
have to be solved at each step in the integration routines.

\begin{figure}[t!p]
\includegraphics[width=0.9\columnwidth]{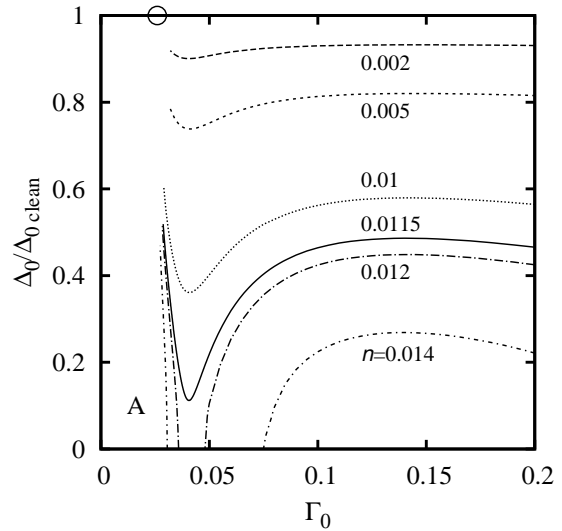}
\vskip -2.3cm
\caption{
The amplitude of the order parameter as a function
of bare hybridization
$\Gamma_0$ for $E_0=-0.1$ and several impurity concentrations.
The approximate location of the critical point where
impurities are decoupled from the conduction band
is marked with a circle.
}
\label{fig2}
\end{figure}

\section{Results}

The calculations were performed for nondegerate impurities,
$N=2$, with superconducting order parameter $\Delta_0 = 0.01$.
All energies are given in units of $D$.

The obtained phase diagram is shown in Fig. 1.
In the Kondo limit the superconducting-normal state
boundary is a straight line. In the mixed valence regime
the boundary for fixed $E_0$ is nonmonotonic.

Fig. 2 shows $\Delta_0$
as a function of $\Gamma_0$ for $E_0=-0.1$.
There is a local gap minimum at intermediate values of $\Gamma_0$, where
$T_K \sim T_{c0}$, where $T_K$ is defined as $\sqrt{\Gamma^2+\epsilon_f^2}$.
Pair breaking is strongest
when the two energy scales are comparable.
For $n \agt 0.0117$ there appears a normal state
around $\Gamma_{0c} \simeq 0.04$.
For $\Gamma_0 < \Gamma_{0c}$, in the region marked ''A''
in Fig. 2 the order parameter quickly rises
to the value of the clean limit. Here the system is very
sensitive to details of interaction. The steep curve
at the boundary of region A means that small change
of $\Gamma_0$ or impurity concentration
may completely alter system properties.
The normal state is separated from the clean-like
superconductor by only about 10\% change of $\Gamma_0$. 
Experiments conducted in this regime may be subject
to greater errors than that in other parts of the phase diagram.
Also the presence of local disorder around each impurity 
significantly influences the system behavior in this limit.
Inhomogeneities in experimental samples,
whether intrinsic to the superconducting state
or influenced by impurity
doping, may lead to situation where the system 
is normal in some regions and superconducting
in the rest of the sample.

\begin{figure}[!tbp]
\includegraphics[width=0.9\columnwidth,clip]{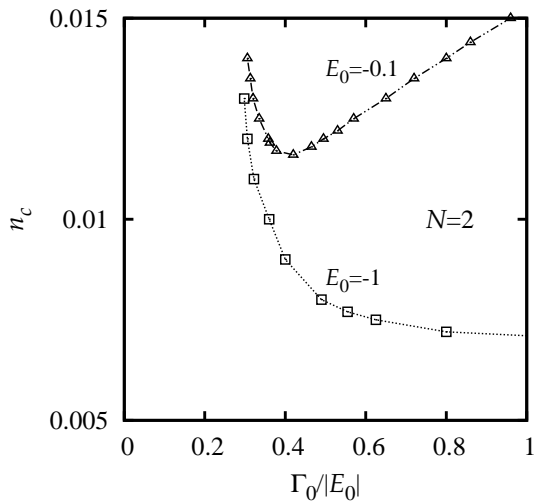}
\vskip -2.5cm
\caption{
Critical concentration for the superconducting-normal transition
as a function of $\Gamma_0=\pi N_0 V^2$
for $E_0=-1$ and $E_0=-0.1$. The $E_0=-1$ curve shows
the behavior in the Kondo limit.
}
\label{fig3}
\end{figure}

For $\Gamma_0 > \Gamma_{0c}$ the impurity resonance
widens and pair breaking at low energies is less effective.
Further increase of $\Gamma_0$ raises pair-breaking
for all energies and eventually the normal state becomes more stable.
Numerical difficulties prevent us from finding the exact
location of the point where
$\Delta_0/\Delta_\textrm{0 clean} \rightarrow 1$.
However this is not necessary for the purpose
of this paper. We can see in Figs. 3 and 4
that decoupling of impurities
from the conduction band occurs at the same
critical value of $\Gamma_0/E_0$ for different
choices of $E_0$.

The phase diagram in the $n$ vs. $\Gamma_0/|E_0|$ plane is shown in Fig. 2.
In the well developed Kondo limit,
when the bare impurity energy level $E_0$
is deep below the Fermi surface
and impurity occupation number $n_f \simeq 1$,
there is only one superconducting-normal transition. This is illustrated
by the $E_0 = -1$ curve in Fig. 3. The transition line
is monotonic in the entire phase diagram. 

When $E_0$ lies closer to the Fermi surface, see curve $E_0=-0.1$,
there exists a critical concentration $n_c^\prime$ such that
for $n < n_c^\prime$ the system remains
superconducting for small and moderate values of $\Gamma_0/|E_0|$
and the transition to the normal state occurs at $\Gamma_0/|E_0| \gg 1$
(beyond the range shown in Fig. 3).
However for $n > n_c^\prime$ there are two additional transitions
for intermediate values of $\Gamma_0/|E_0|$, when
$T_K \sim T_{c0}$.

\begin{figure}[!tbp]
\includegraphics[width=0.9\columnwidth,clip]{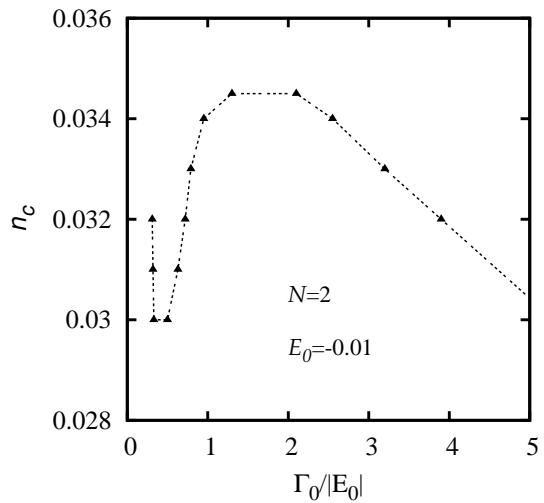}
\vskip -2.5cm
\caption{The phase diagram in the mixed valence limit,
$E_0 = -0.01$. The scale of the horizontal axis is changed
relative to Fig. 2 to illustrate behavior at large $\Gamma_0$.
}
\label{fig4}
\end{figure}

This is clearly shown in Fig. 4 where $E_0=-0.01$.
The horizontal
axis of Fig. 4 is extended to larger values of $\Gamma_0$.
This is the general form of the phase diagram when $E_0$
is close to the Fermi energy.
The superconducting state
boundary for $E_0=-0.1$ has similar shape
(not shown in Fig. 3 due to smaller horizontal scale).
For increasing $E_0$ the phase transition line
is shifted towards higher impurity concentrations
and the normal state section at intermediate $\Gamma_0 /|E_0|$
becomes narrower. This can be explained by the weaker
pair breaking in the mixed valence regime.

The impurity-induced peak in the conduction electron DOS
remains at $\omega=0$ for all $E_0 < 0$. The peak splits
in two, one at positive $\omega$ and one at negative
$\omega$ when $\epsilon_f \gg \Gamma$. For a nondegenerate
impurity in a superconductor with order parameter
having lines of nodes this occurs only at $E_0 > 0$.
This conclusion was verified numerically.

\begin{figure}[!tbp]
\includegraphics[width=0.9\columnwidth,clip]
{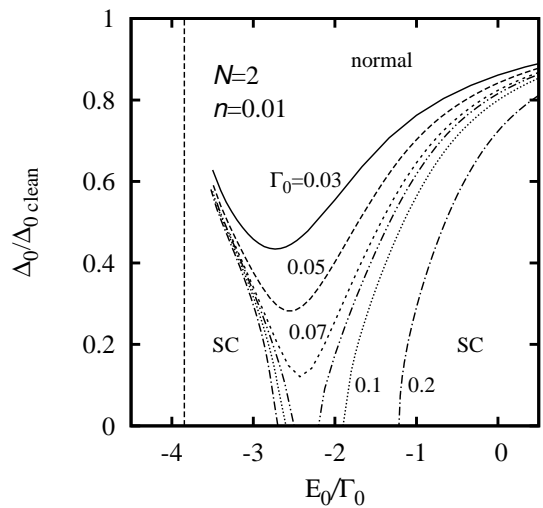}
\vskip -2.6cm
\caption{Superconducting order parameter
$\Delta_0$ as a function of $E_0$
for several values of $\Gamma_0$ at $n=0.01$.
The unmarked dash-dotted curve is the result
for $\Gamma_0=0.08$. The vertical dotted line
marks the impurity quantum phase transition.
}
\label{fig5}
\end{figure}

\vfil\eject
The dependence of the order parameter on $E_0$ for
several values of $\Gamma_0$ is shown in Fig. \ref{fig5}.
This behavior is always nonmonotonic. The minimum
$\Delta_0$ occurs when $T_K \sim \Delta_0$.
For larger $\Gamma_0$ we have two superconducting regions,
one in the Kondo limit and one in the mixed valence regime,
separated by the normal state.

\section{Conclusions}

The impurity quantum phase transition at finite coupling
is associated with the particle-hole asymmetry of the model.\cite{ingersent98}
If the impurity degeneracy $N$ is increased, the impurity transition
occurs at lower coupling.\cite{lsb2008} This is expected since
for larger $N$ the impurity resonance is located at higher energy.
The precise location of the main features
of the phase diagram changes but its overall qualitative
form remains intact.\cite{lsb2008}

The suppression of the superconducting state
along the diagonal of the phase diagram in Fig. 1
results from the interplay of energy scales
and is not directly related to the symmetry
of the order parameter.
In an s-wave superconductor the steepest initial decrease
of the superconducting critical temperature
as a function of impurity concentration, $(dT_{c}/dn)_{|{T_c=T_{c0}}}$
occurs for $T_{c0} \sim T_K$.\cite{bickers87}
At $T=0$ this corresponds to $\Delta_0 \sim T_K$
and we expect a normal state insertion similar to that in Fig. 1
also for order parameters without nodes on the Fermi surface.

Features of the phase diagram
should be visible in experiments by (a) changing impurity
concentration at fixed carrier doping
and (b) varying the strength of coupling
to the impurity, the ratio $\Gamma_0/E_0$, at fixed
impurity concentration. 

The depth of the impurity level
may be tuned by applying
pressure. There exist studies of heavy-fermion compounds,
most notably $\rm CeCu_2(Si_{1-x}Ge_x)_2$
where varying hydrostatic pressure reveals two superconducting
regions in the phase diagram.\cite{steglich2004}.
The change of pressure shifts the chemical potential. Position
of the bare impurity level $E_0$ of Ce ions
relative to $E_F$ changes accordingly.
This may explain the existence of two superconducting regions
in the phase diagram of $\rm CeCu_2(Si_{1-x}Ge_x)_2$
similarly to results of this work.
Increasing pressure brings the system into
the mixed-valence regime. It is equivalent
to increasing the ratio $\Gamma_0/|E_0|$ in our work.
The existing explanation for the superconducting region
at high pressure uses the concept of the valence-fluctuation
mediated pairing mechanism.\cite{miyake2002}
It is interesting to note that a qualitatively
similar nonmonotonic behavior
of superconducting critical temperature 
is obtained in an impurity problem when increasing $E_0$.
We hope that similar experiments under varying pressure
may be conducted
on superconductors with Anderson impurities.
They should reveal the details of the phase diagram
described in this article.


A nontrivial question is realistic treatment
of finite impurity concentrations in this problem.
In real samples there are local inhomogeneities
and variations of impurity concentration.
There is some evidence that inhomogeneity
of the superconducting state may
be intrinsic in some high-$T_c$ compounds
even in absence of impurities in $\rm CuO_2$
planes.\cite{hirschfeld2008} This may have important
consequences near the transition line
in the region A of the phase diagram in Fig. 2, where
a small change in hybridization or impurity level $E_0$
implies a large change in properties.
In this limit the entire sample might even consist of a mixture
of normal and superconducting patches.
This sensitivity should be also visible
in the low-energy physics
near the impurity critical point.
In inhomogeneous samples pockets of clean-like
superconductor may coexist with regions
where low-energy physics is dominated by Kondo
impurities. Scanning tunneling microscopy (STM)
 measurements should be helpful
in investigating the phase diagram in the vicinity
of the superconducting state boundary
and the impurity critical point.



\end{document}